\documentclass[prb,aps,twocolumn,footinbib,floatfix,10pt,longbibliography,superscriptaddress]{revtex4-1}

\usepackage{comment}
\usepackage{chemformula} 
\usepackage[T1]{fontenc} 
\usepackage{graphicx}
\usepackage{dcolumn}
\usepackage{lipsum}
\usepackage{tabu} 
\usepackage{mathtools}

\usepackage{braket}
\usepackage{color}
\usepackage[11pt]{moresize}
\usepackage{anyfontsize}
\usepackage{bbding}
\usepackage{amsmath}
\usepackage{xcolor}
\usepackage{braket}
\usepackage[normalem]{ulem}
\usepackage{gensymb}

\newcommand{\SiN}{Si$_3$N$_4$}
\newcommand{\SiO}{SiO$_2$}
\newcommand{\Psat}{$P_{\text{sat}}$}
\begin{document} 

\title{Advanced architectures for coupling III-V nanowires to photonic integrated circuitry}

\author{Edith Yeung}
\affiliation{University of Ottawa, Ottawa, Ontario, Canada, K1N 6N5.}
\affiliation{National Research Council Canada, Ottawa, Ontario, Canada, K1A 0R6.}
\author{Kataryna Sorensen}
\affiliation{University of Ottawa, Ottawa, Ontario, Canada, K1N 6N5.}
\affiliation{National Research Council Canada, Ottawa, Ontario, Canada, K1A 0R6.}
\author{David B. Northeast}
\affiliation{National Research Council Canada, Ottawa, Ontario, Canada, K1A 0R6.}
\author{Maziyar Milanizadeh}
\affiliation{National Research Council Canada, Ottawa, Ontario, Canada, K1A 0R6.}
\author{Philip J. Poole}
\affiliation{National Research Council Canada, Ottawa, Ontario, Canada, K1A 0R6.}
\author{Robin L. Williams}
\affiliation{National Research Council Canada, Ottawa, Ontario, Canada, K1A 0R6.}
\author{Dan Dalacu}
\affiliation{University of Ottawa, Ottawa, Ontario, Canada, K1N 6N5.}
\affiliation{National Research Council Canada, Ottawa, Ontario, Canada, K1A 0R6.}

\begin{abstract}
This work implements a hybrid device based on a semiconductor quantum dot embedded within a nanowire to bridge a non-continuous curved waveguide structure. The geometry takes advantage of evanescent coupling between the photonic structures to recover single photons emitted from both outputs of the device. Auto- and cross-correlation measurements were performed on different output facets of the device. We demonstrate single-photon emission from both ends of the nanowire for both neutral, $X$ and $XX$, and charged $X^-$, excitonic complexes. We further demonstrate the cascaded $XX-X$ emission by collecting each complex from a different facet. This work lays the foundation for on-chip architectures which utilize multi-directional integration of quantum emitters.

\end{abstract}

\vspace{0.5cm}

\vspace{0.5cm}

\maketitle 

\section{Introduction}

Sources of single-photons are a key resource in quantum information processing applications\cite{Obrien_NP2009} and solid-state two-level systems\cite{Aharonovich_NP2016} offer a promising route to generating single-photons on-demand with high efficiency. Semiconductor quantum dots (QDs) are of particular relevance as they have achieved nearly ideal performance when integrated with photonic structures\cite{Senellart_NN2017}. These structures, however, are typically designed to operate out-of-plane whereas future applications will likely require the scalability provided by on-chip integration\cite{Hepp_AQT2019}.

QDs embedded within nanowires that are grown\cite{Dalacu_APL2011} (i.e. bottom-up) as opposed to etched\cite{Claudon_NP2010} (i.e. top-down) provide a route to fabricating sources of single photons deterministically. The number of emitters per device is controlled and each device is grown at a specified position on the substrate. Bottom-up nanowire QDs have been used to generate both high quality single photons\cite{Dalacu_NL2012,Laferriere_SR2022} and high fidelity entangled photon pairs~\cite{Versteegh_NC2014,Pennacchietti_CP2024}. These sources were also operated out-of-plane and high efficiency was obtained by designing the nanowire to act as a single mode waveguide supporting only the fundamental HE$_{11}$ mode\cite{Dalacu_NT2019}, and by introducing a taper along the length of the nanowire\cite{Dalacu_NANOM2021} to expand the mode for efficient coupling to external optics.  Properly designed, 95\% of the dot emission is directed into the HE$_{11}$ mode\cite{Dalacu_NT2019} which can be coupled to single mode fibre with an efficiency of 93\%\cite{Bulgarini_NL2014}. 

\begin{figure}
    \centering
    \includegraphics[width=\linewidth]{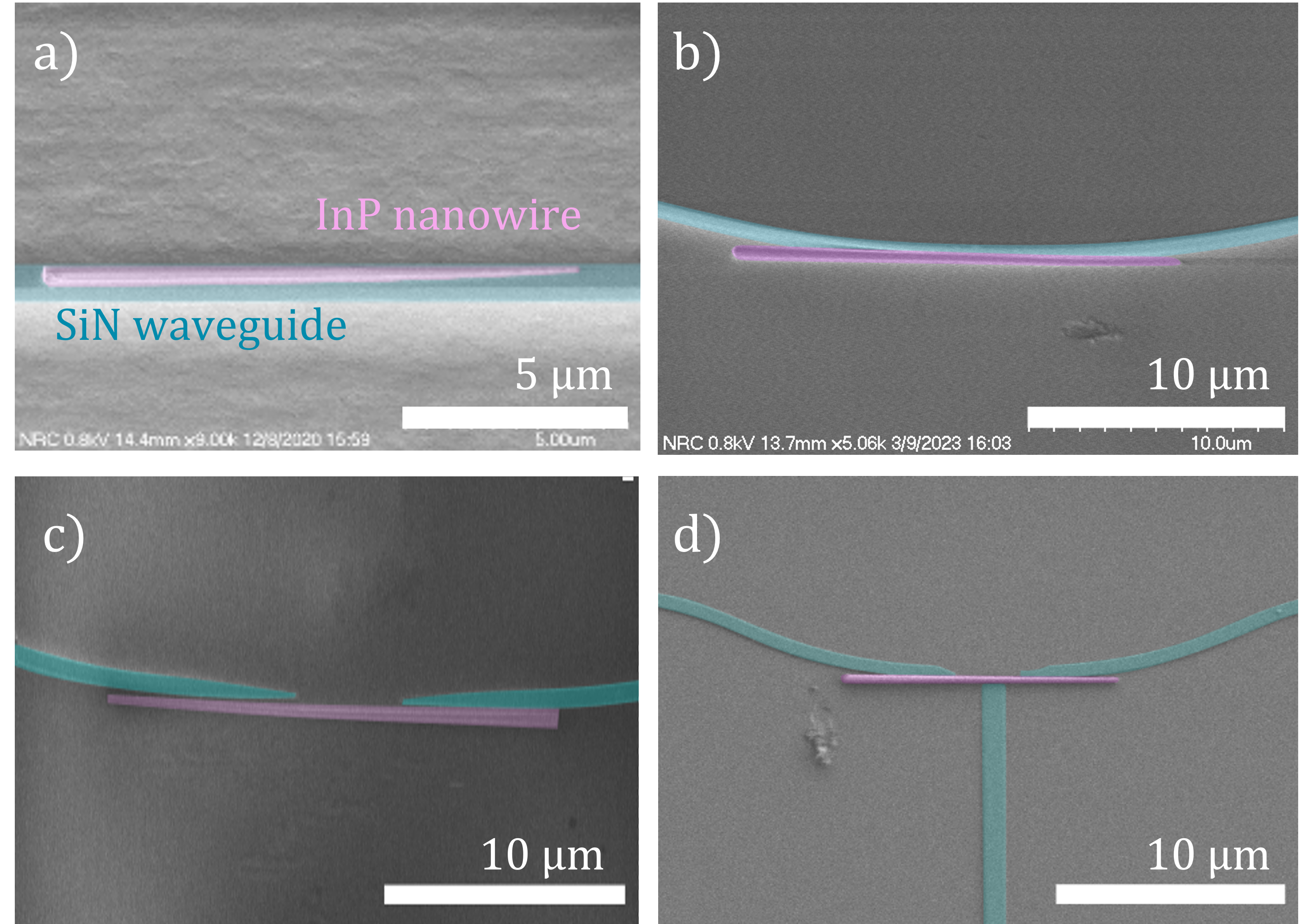}
    \caption{False colour SEM images with \SiO\ in grey, \SiN\ in blue and InP in purple. (a) Straight \SiN\ waveguide with a tapered InP nanowire on top. (b) Curved waveguide with a nanowire on the side. (c) Segmented waveguide with a nanowire bridging the gap. (d) As in (c) but with an additional waveguide for cross-pumping.}
    \label{fig:architectures}
\end{figure}

The ability to control the nanowire taper is also relevant for designing chip-integrated sources. In this case, the mode expansion provided by the taper is not for efficient coupling to low numerical aperture optics, but rather for efficient evanescent coupling to on-chip waveguides. By placing an appropriately tapered nanowire along a straight \SiN\ waveguide, as in Fig.~\ref{fig:architectures}(a), the HE$_{11}$ mode can be efficiently coupled on-chip, with simulated coupling efficiencies approaching 100\% \cite{Mnaymneh_AQT2020,Yeung_PRB2023}. The hybrid structures can be assembled by picking up nanowires from the growth substrate and placing them on \SiN\ waveguides using a scanning electron microscope (SEM)-based nanomanipulator\cite{Mnaymneh_AQT2020}. With this approach, we have demonstrated >90\% coupling of the HE$_{11}$ mode to a \SiN\ waveguide and have generated on-chip single photons with purities exceeding 95\%\cite{Yeung_PRB2023}. 

The pick and place hybrid platform is not limited to simple structures such as that shown in Fig.~\ref{fig:architectures}(a): one can also construct hybrid devices consisting of a nominally untapered nanowire with a thinner diameter placed beside a curved waveguide as shown in Fig.~\ref{fig:architectures}(b). In this case the quantum dot within the nanowire is located at its midpoint and the evanescent transfer of the nanowire mode to the waveguide occurs as the separation between the two increases. Such a device can harvest photons emitted in either direction along the nanowire waveguide, a limitation of the unidirectional structure in Fig.~\ref{fig:architectures}(a) where emission towards the base is not efficiently collected.  

Expanding on the curved waveguide architecture, a gap can be introduced in the waveguide with the nanowire connecting the circuit, as shown in Fig.~\ref{fig:architectures}(c). In this case, the waveguide mode can be transferred completely to the nanowire and back again, assisted by tapering each end of the segmented waveguide. This structure is relevant for new types of experiments such as transmission measurements to investigate coherent interactions between single photons and two-level systems\cite{Thyrrestrup_NL2018}. By incorporating an additional waveguide orthogonal to the nanowire, as in Fig.~\ref{fig:architectures}(d), we can also envision a resonant fluorescent excitation scheme where pump laser rejection is achieved using orthogonal excitation and collection directions \cite{Ates_PRL2009,Flagg_NP2009,Muller_PRL2007}. 

In this article we explore some of the advanced architectures shown in Fig.~\ref{fig:architectures} for on-chip single photon generation in Si-based integrated photonic circuitry.  We evaluate, through simulations, geometry-dependent coupling efficiencies between a nanowire and a curved waveguide, both with and without a gap. We find the segmented waveguide design more robust against variations in the nanowire diameter and use this architecture to demonstrate efficient generation of on-chip high purity single photons. We further utilize the bi-directional nature of the device to perform cross-correlation measurements by collecting from both ends of the nanowire with the nanowire acting as an integrated beamsplitter.

\section{Simulations}

As mentioned in the introduction, a nanowire containing a quantum emitter can transfer its light efficiently to a nearby waveguide by evanescent coupling~\cite{Mnaymneh_AQT2020,Yeung_PRB2023}. This process is robust provided both structures allow for a single mode and a tapered section is included over sufficient length to allow adiabatic transfer of mode energy between the two. In this scenario, light that is emitted away from the taper—50\% of light from the quantum dot—will encounter an abrupt end of the nanowire and will couple poorly into the waveguide. A waveguide that curves as it approaches the nanowire will be seen as a gradual change in the refractive index. Device designs can use this effect to allow for light coupling in both directions. Since quantum dot emission has a narrow bandwidth and optical response of all structures is expected to change slowly with wavelength, frequency domain finite element simulations (COMSOL Multiphysics) are performed to predict device performance.

\begin{figure}[h]
    \centering
    \includegraphics[width=0.9\linewidth]{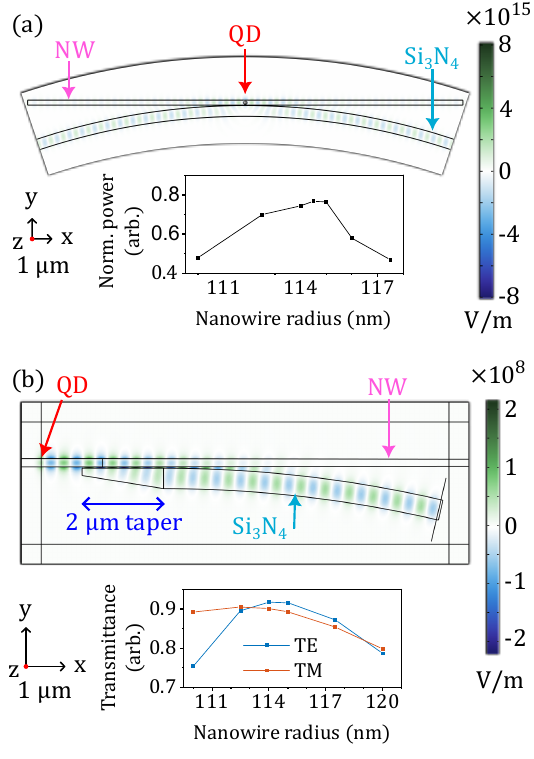}
    \caption{Finite element simulations were performed for the two different hybrid coupling approaches. (a) Top-down view of the photonic integrated circuit. The quantum dot emitter (here a dipole) is located at the midpoint of the InP nanowire (NW) which is tangent to the curved \SiN\ waveguide. The emission from the quantum dot is emitted into the nanowire and couples into the waveguide as it bends away. The total power coupled into the waveguide as a function of nanowire radius, $r_{\text{nw}}$, is shown in the inset. (b) Revised coupling design that makes use of a gap in the \SiN\ waveguide with tapered sections at the gap. The inset shows the improved transmittance between both HE$_{11}$ polarizations in the nanowire and the TE/TM modes in the \SiN\ ridge waveguide.}
    \label{fig:sim_plots}
\end{figure}

We first investigate power transfer from a dipole in a straight InP nanowire to a curved \SiN\ waveguide. A 20\,$\mu$m nanowire length is used in order to provide sufficient separation between the nanowire ends and the curved waveguide for a waveguide with a 30\,$\mu$m radius of curvature. The \SiN\ waveguides are 0.4\,$\mu$m thick and 0.5\,$\mu$m wide, chosen to support a single mode in both TE and TM polarizations, sitting on 4\,$\mu$m of SiO$_2$. The nanowire, with radius $r_{\text{nw}}$, is placed as seen in Fig.~\ref{fig:sim_plots}(a) such that the dipole is located precisely where the nanowire meets the waveguide. The figure shows the electric field polarized along the TE direction of the waveguide. In the inset, we show the fraction of the total power radiated from the dipole that couples to the waveguide for different $r_{\text{nw}}$. We calculate a peak coupling of greater than 76\% in both waveguide directions which drops quickly for nanowire radii away for the optimal value of $r_{\text{nw}} = 115$\,nm. 

Combining the curved geometry with a tapered waveguide, power transfer can be made less sensitive to the nanowire radius, $r_{\text{nw}}$. Fig.~\ref{fig:sim_plots}(b) depicts a curved \SiN\ waveguide that tapers to an end over a length of 2\,$\mu$m. Simulated transmittance of HE$_{11}$ nanowire modes to TE and TM waveguide modes is shown in the inset for different $r_{\text{nw}}$. We calculate a peak transmission of over 90\% for both linear polarizations and observe a reduced dependence on $r_{\text{nw}}$ compared to the untapered waveguide geometry in Fig.~\ref{fig:sim_plots}(a). In the  experiments below, we will focus on the segmented tapered \SiN\ geometry as it provides higher coupling values with a higher tolerance for variations in nanowire radius.

\section{Experimental Details}

The quantum emitters used in this work are InAsP quantum dots embedded within InP nanowires. An array of position-controlled InP nanowires are grown using selective-area vapour-liquid-solid epitaxy, see Refs.~\citenum{Dalacu_APL2011,Dalacu_NL2012,Laferriere_SR2022} for details. In the first growth stage, we grow a single InAsP QD as a segment within a 20\,nm diameter InP nanowire core at a nominal height of 10\,$\mu$m above the substrate. In the second growth stage, the nanowire core is clad with InP to increase the diameter to 210\,nm and to increase the nanowire height to 20\,$\mu$m such that the QD is at the midpoint of the nanowire. In this cladding stage, growth conditions are adjusted to minimize tapering of the nanowire diameter along its length as the proposed designed is based on an untapered nanowire, see  Fig.~\ref{fig:sim_plots}.

The integrated photonic circuitry is based on low-pressure chemical vapour deposited \SiN\ waveguides fabricated on Si wafers using a commercial foundry. Waveguide dimensions are 0.4\,$\mu$m tall and 0.5\,$\mu$m wide clad with 3.3\,$\mu$m of \SiO\ above and 4\,$\mu$m below.  For coupling on and off chip, we use edge couplers designed for lensed fibres: at the etched facets, the waveguide widths are tapered to 0.2\,$\mu$m over a length 150\,$\mu$m. The segmented waveguide consists of a curved section with a 30\,$\mu$m radius of curvature and with a 4\,$\mu$m gap at the midpoint. On either side of the gap, the waveguide is tapered down to 0.180\,$\mu$m over 2.0\,$\mu$m, as shown in Fig.~\ref{fig:sim_plots}(b). A window in the top \SiO\ layer is opened to expose just the segmented waveguide section, permitting nanowire placement. The pre-characterized nanowire is picked up from the growth substrate with a piezo-motor controlled tungsten tip on the nanomanipulator inside an SEM \cite{Mnaymneh_AQT2020,Yeung_PRB2023}. The nanowire is placed tangential to the curved waveguide such that it bridges the gap with the midpoint of the nanowire, and hence the quantum dot, centred in the gap. The final structure, which we term a gap-coupled device, is shown in Fig.~\ref{fig:architectures}(c).

The device was measured in a fibre-coupled closed-cycle helium cryostat at 5.5\,K. The cyrostat is equipped with three xyz piezo stacks: one for free-space alignment and two for aligning fibres to edge facet couplers. We take advantage of the on-chip platform to pump the quantum dot and collect the emission in three different configurations: ($i$) pump via fibre through the waveguide from one side and collect from the other side, ($ii$) pump and collect from the same side using a 90:10 fibre splitter (here we can also simultaneously collect from the other side), and ($iii$) pump from the top using free-space optics and collect from both sides of the segemented waveguide.   

Photoluminescence (PL) measurements were done using above-band continuous wave (CW) or pulsed laser  excitation. The emission was collected using one of the above configurations and sent to a fibre-coupled grating spectrometer for spectrally-resolved measurements or to a fibre-coupled narrow-band (bandwidth $BW = 0.1$\,nm) tuneable filter for measurements on single excitonic lines. For measurements on the single excitonic complexes, including time-resolved PL, high-resolution PL and time-correlated PL, emitted photons were detected using superconducting nanowire single-photon detectors (SNSPDs) with efficiencies of 80\% and timing jitters of 100\,ps. 

\section{Results}

\begin{figure}
    \centering
    \includegraphics[width=0.9\linewidth]{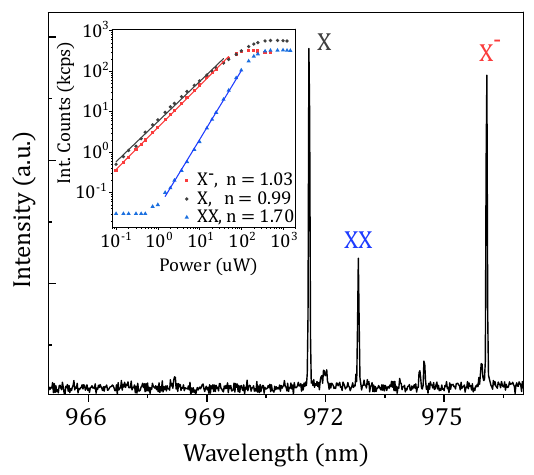}
    \caption{
    PL spectrum measured on a gap-coupled device showing three dominant peaks identified as the excitonic complexes $X$, $XX$ and $X^-$. The inset shows the integrated intensity as a function of pulsed excitation power at 80\,MHz for each complex: linear ($\mathrm{n}\sim1$) for $X$ and $X^-$ and approaching quadratic ($\mathrm{n}\rightarrow 2$) for $XX$ \cite{Sek_JAP2010}.}
    \label{fig:pl}
\end{figure}

In Fig.~\ref{fig:pl} we show a PL spectrum from the gap-coupled device shown in Fig.~\ref{fig:architectures}(c). The measurement was performed by exciting through the waveguide from one side and collecting from the other side. The spectrum shows three dominant peaks, typical in this quantum dot system (see, for example, Ref.~\citenum{Laferriere_APL2021}). The peaks are identified as emission from the neutral exciton ($X$), biexciton ($XX$) and trion ($X^-$) where peak identification is based on the intensity dependence on pump power (see inset) as well as high-resolution PL and cross-correlation measurements discussed below.

The integrated intensities from each complex for pulsed excitation at 80\,MHz at saturation, $P_{\mathrm{sat}}$, were 575\,kcps for $X$, 330\,kcps for $XX$ and 330\,kcps from $X^-$.  Using these values, we obtain end-to-end efficiencies (detected counts/80\,MHz) of 0.72\%, 0.41\%, and 0.41\% for $X$, $XX$ and $X^-$, respectively. These efficiencies are limited predominantly by the throughput of the setup ($<10\%$) and by the loss of 50\% of the photons directed towards the pump direction as here we are only collecting from one side. Additional losses include the nanowire to waveguide coupling and the detector efficiency. If we reverse the excitation and collection waveguides, we observe a decrease in the count rate of two orders of magnitude whereas simulations predict equal coupling in both directions. This asymmetry may be due to a misplacement of the QD in the nanowire with respect to the gap in the waveguide. A direction-dependent coupling may also arise from an unintentional taper of the nanowire that can be observed in Fig.~\ref{fig:architectures}(c). Nonetheless, we obtain sufficient counts from both directions to perform the bi-directional measurements that will be discussed below.

\begin{figure}
    \centering
    \includegraphics[width=\linewidth]{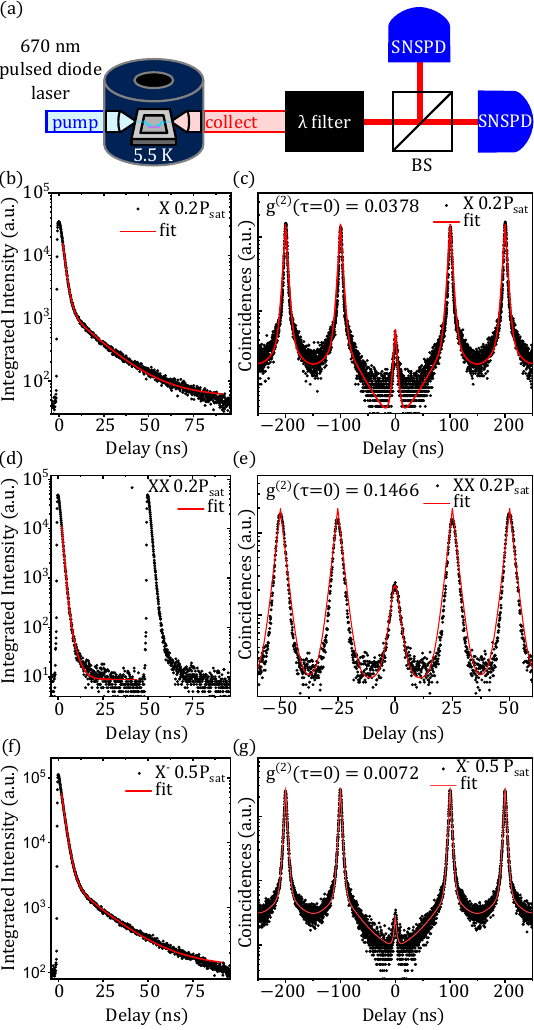}
    \caption{(a) Schematic of the setup for measuring $g^{(2)}(\tau)$. The excitonic complex to be measured is first filtered then directed to a beamsplitter with the output ports sent to two SNSPD detectors. (b, d, f) PL decay traces and (c, e, g) auto-correlation measurement for the three complexes, $X$, $XX$, and $X^-$, respectively, plotted semi-logarithmically. The excitation rate used in each measurement is shown in the figure as a fraction of the saturation power, $P_{\text{sat}}$.}
    \label{fig:trad_g2}
\end{figure}

Time-resolved PL traces of the three complexes are shown in Fig.~\ref{fig:trad_g2}(b,d,f) where we excite from one side and collect for the other as above, see Fig.~\ref{fig:trad_g2}(a). To extract decay rates, we assume a double exponential of the form \cite{Johansen_PRB2010}
\begin{equation}
    I(\tau)=A_be^{-\frac{\tau}{\tau_b}}+A_de^{-\frac{\tau}{\tau_d}}+y_0
\end{equation} 
where $\tau$ is the delay between a synchronization pulse from the laser and a detection event, $1/\tau_b$ is the radiative decay rate of the bright exciton, $1/\tau_d$ is the spin-flip rate that converts the system from a dark to bright state with $A_b$ and $A_d$ the amplitudes of the bright and dark contributions, respectively, and $y_0$ is a background offset. The ratio $A_b/A_d$ is a measure of the occupation probability of the bright state. Model fits to the decay curves are shown as red lines in Fig.~\ref{fig:trad_g2}(b,d,f) and extracted parameters $\tau_b$, $\tau_d$ and $A_b/A_d$ are summarized in Table.~\ref{tab:lifetimes} for all three complexes.

The time-resolved PL traces of the $X$ and $X^-$ emission, Fig.~\ref{fig:trad_g2}(b) and (f), respectively, both show strong slow decay contributions with decay times, $\tau_d$, consistent with spin-flip rates of neutral excitons in quantum dot systems\cite{Johansen_PRB2010}. For $X^-$, the spin-flip is associated with a carrier in a p-shell of the dot: a dark excited state of the complex that relaxes to the bright state\cite{Laferriere_SR2022}. In contrast, the PL decay of the $XX$ emission, Fig.~\ref{fig:trad_g2}(d), is predominantly mono-exponential with a possible small contribution from a slightly slower process. 

Second-order autocorrelations, $g^{(2)}(\tau)$, measured using pulsed diode excitation ($\lambda = 670$\,nm, pulse width $=100$\,ps) in a Hanbury-Brown and Twiss (HBT) setup, see Fig.~\ref{fig:trad_g2}(a), are shown in Fig.~\ref{fig:trad_g2}(c,e,g) for $X$, $XX$ and $X^-$, respectively. Due to the significant slow component in the decay of $X$ and $X^-$, an excitation repetition rate of 10\,MHz was used to allow the system to fully decay between pulses. For $XX$, where $\tau_d$ was significantly shorter, the dot was excited at 40\,MHz. The reduced  zero-delay peak in the $g^{(2)}(\tau)$ for all three complexes demonstrates single photon emission. To quantify the single photon purity, the $g^{(2)}(\tau)$ is modeled by
\begin{equation}
\begin{split}
    g^{(2)}(\tau)= &C_0\left[e^{-\frac{|\tau|}{\tau_b}}+\frac{A_d}{A_b}e^{-\frac{|\tau|}{\tau_d}}\right]
                    -B_0e^{-\frac{|\tau|}{\tau_{\text{re}}}} \\
                    &+C_1\sum_{n\neq0}\left[e^{-\frac{|\tau|-nt_{\text{rep}}}{\tau_b}} +\frac{A_d}{A_b}e^{-\frac{|\tau|-nt_{\text{rep}}}{\tau_d}}\right]\\
                    &+C_b
\end{split}
\label{eqn_2}
\end{equation}where $C_0$ is the amplitude of the coincidences due to the multiphoton emission events at $\tau=0$, $B_0$ is associated with re-excitation of the dot, at a rate of $1/\tau_{\text{re}}$, from the same excitation pulse, $C_1$ corresponds to coincidences recorded from different excitation pulses at $\tau=nt_{\text{rep}}$ for $n\neq0$, and $C_b$ represents the background counts. For the remaining parameters, values extracted from the fits to the time-resoled PL are used. The model fits are shown in Fig.~\ref{fig:trad_g2}(c, e, g) (red curves) for the three complexes. From the ratio of the area of the zero delay peak to the average peak area at $\tau=nt_{\text{rep}}$, we calculate the probability of multiphoton emission, $g^{(2)}(\tau=0)$, listed in Table.~\ref{tab:lifetimes}.

\begin{table}[]
    \centering
    \begin{tabular}{c|c|c|c|c}
         Complex &  $\tau_b$ (ns) & $\tau_d$ (ns) & $A_d/A_b$ & $g^{(2)}(\tau=0)$\\ [0.5ex] 
         \hline
         $X$ & $1.680\pm0.004$& $16.2\pm0.2$& 0.0296&0.03776\\
         $XX$ & $0.982 \pm 0.006$ & $2.8 \pm 0.3$ & 0.011 & 0.14663\\
         $X^-$ & $2.107 \pm 0.002$&$18.2 \pm 0.2$&0.046& 0.00724
    \end{tabular}
    \caption{Summary of parameters extracted from model fits to the PL decay traces and second-order correlation measurements.}
    \label{tab:lifetimes}
\end{table}

We obtain very low probabilities of multi-photon emission from $X$ and $X^-$ with $g^{(2)}(\tau=0)$ values of $<4\%$ and $<1\%$, respectively. Coincidences in the zero-delay peak are associated with re-excitation of the dot from the same laser pulse, previously seen in this system with above-band excitation~\cite{Laferriere_NL2020}. We note that the dip in the zero-delay peak that is characteristic of re-excitation is not observed and we speculate that the timescales associated with the re-excitation process are too fast to be resolved using our 100\,ps detectors. The higher $g^{(2)}(\tau=0)$ for $XX$ of nearly 15\% may be related to the faster decay ($\tau_b = 982$\,ps) and/or higher excitation power required for saturation, both of which would lead to an increased re-excitation probability during the 100\,ps excitation pulse. 

\begin{figure}
    \centering
    \includegraphics[width=\linewidth]{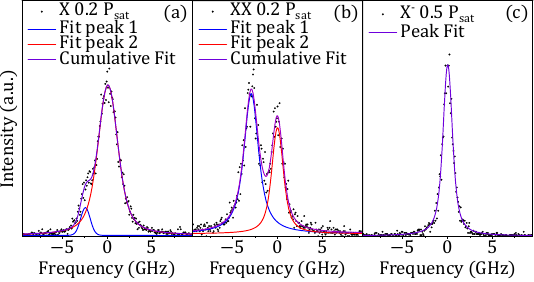}
    \caption{High-resolution PL of the (a) $X$, (b) $XX$ and (c) $X^-$ peaks measured at excitation powers indicated. Model fits are shown in purple. For the neutral complexes, the two peaks that constitute the doublets are shown in blue and red.}
    \label{fig:linewidths}
\end{figure}

To evaluate the spectral purity of the emitted photons, we measure high-resolution PL spectra for each complex by directing the filtered emission through a tuneable Fabry-Perot etalon with a bandwidth of $BW= 200$\,MHz, sufficiently narrow to resolve real lineshapes. High-resolution spectra of $X$ and $XX$ are shown in Fig.~\ref{fig:linewidths}(a) and (b), respectively. In each case, two peaks are resolved corresponding to the two orthogonal exciton states separated by the fine structure spitting, $FSS$\cite{Bayer_PRB2002}. The peaks are fit with Voigt lineshapes from which we extract an $FSS = 2.55\pm0.02$\,GHz for $X$ and $2.96\pm0.01$\,GHz for $XX$. These values of $FSS$ are larger then we typically observe in this system\cite{Versteegh_NC2014} which we associated with slower switching between Group V precursors during the dot growth in this sample that results in an axial asymmetry due to P and As tailing\cite{Cygorek_PRB2020}.

The extracted linewidths are $\delta\omega_{V,X_1}=1.83\pm0.07$\,GHz and $\delta\omega_{V,X_2}=2.59\pm0.03$\,GHz for the two $X$ dipoles and $\delta\omega_{V,XX_1}=1.85\pm0.05$\,GHz and $\delta\omega_{V,XX_2}=1.66\pm0.05$\,GHz for the two $XX$ dipoles. In an ideal system, the homogeneous broadening of the linewidth is limited by its lifetime giving a Lorentzian lineshape with a transform-limited linewidth of $\delta\omega=1/2\pi \tau_b$ \cite{Laferriere_PRB2023}. We observe Voigt lineshapes indicating a Gaussian component (i.e. inhomogeneous broadening) likely from a fluctuating charge environment\cite{Robinson_PRB2000} common with above-band excitation. Using measured lifetimes, we calculate transform-limited linewidths of $1/2\pi \tau_{b,X}=0.0947$ GHz for $X$ and $1/2\pi \tau_{b,XX}=0.162$\,GHz for $XX$. Observed linewidths are 19.3 and 27.3 (11.4 and 10.2) times this limit for $X$ ($XX$), depending on the dipole, which gives a measure of the charge noise. The reduced excess broadening observed for $XX$ may indicate a reduced sensitivity to charge fluctuations for complexes with beginning and end states both modified by the presence of charge; for $X$, where the final state in empty, this is not the case. 

For the $X^-$ transition, shown in Fig.~\ref{fig:linewidths}(c), a single peak is observed, consistent with its identification as a singly charged complex and thus not expected to have a $FSS$\cite{Bayer_PRB2002}. The lineshape here is also Voigt with a linewidth of $\delta\omega_{V,X^-}=1.311\pm0.007$\,GHz, 17.4 times the transform-limited linewidth of $1/2\pi \tau_{b,X^-}=0.0755$\,GHz. We note that above-band excitation is not expected to produce lifetime-limited photons\cite{Kuhlmann_NC2015}, however, the nanowire quantum dot system has been shown to approach this limit\cite{Laferriere_PRB2023}. It is possible that the slower Group V switching in this sample may have influenced the charge environment, resulting in the observed levels of excess broadening.

\begin{figure}
    \centering    \includegraphics[width=\linewidth]{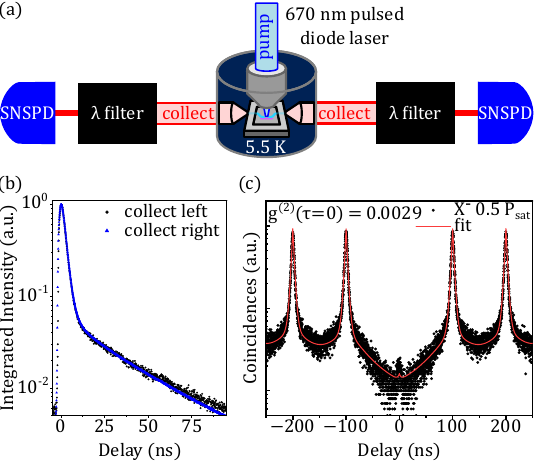}
    \caption{(a) Schematic of bi-directional collection from the gap-coupled device. (b) Comparison of the PL decay of $X^-$ emission for collection from the left and right facets of the chip. (c) Coincidences, $g^{(2)}(\tau)$, measured using the configuration in (a). For each measurement, the excitation power was 0.5\Psat.}
    \label{fig:bi_g2}
\end{figure}

Measurements thus far have demonstrated efficient coupling of single photons on chip, similar to previous experiments using straight waveguides\cite{Mnaymneh_AQT2020,Yeung_PRB2023}. Next we harness the versatility of the gap-coupled architecture to perform bi-directional measurements. In this case, we excite the nanowire from above using free-space optics and collect from both edge facets, as shown in Fig.~\ref{fig:bi_g2}(a). Fig.~\ref{fig:bi_g2}(b) compares time-resolved PL traces of the $X^-$ emission for collection from the left and right facets. The two traces lie on top of each other with decay times equivalent to that obtained when excitation was via a waveguide, see Fig.~\ref{fig:trad_g2}(f). This measurement simply confirms that the excitation and collection configuration does not impact the decay processes of the transition.

Next we perform a second-order correlation measurement where collection from both ends of the nanowire replaces the beamsplitter in the HBT set-up (compare Fig.~\ref{fig:trad_g2}(a)). The $g^{(2)}(\tau)$ measured at an excitation power of 0.5\Psat\ is shown in Fig~\ref{fig:bi_g2}(c). Using Eq.~\ref{eqn_2}, we obtain $g^{(2)}(\tau=0)\sim0.003$, increasing to $g^{(2)}(\tau=0)\sim0.013$ at \Psat. The low level of coincidences in the zero delay peak confirms, again, the single photon nature of the emission from the quantum dot.

\begin{figure}
    \centering
    \includegraphics[width=\linewidth]{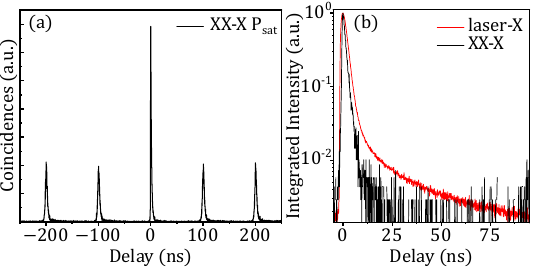}
    \caption{(a) $XX$-$X$ cross-correlation measured using bi-directional collection at $P_{\mathrm{sat}}$. (b) Zoom in showing PL decay of the $X$ emission triggered using $XX$ detection (black curve) and a synchronization pulse from the laser (red curve).}
    \label{fig:g2_XX_X}
\end{figure}

Using the same configuration, we can also perform cross-correlation measurements by filtering for different complexes at each output. Fig.~\ref{fig:g2_XX_X}(a) shows the cross-correlation between $XX$ and $X$ photons where we collect $X$ photons from one side of the chip and $XX$ photons from the other. With $XX$ sent to the 'start' detector and $X$ to the 'stop' detector we observe strong bunching for $\tau>0$ consistent with the cascaded $XX-X$ emission process \cite{Versteegh_NC2014}.

In Fig.~\ref{fig:g2_XX_X}(b), we zoom in on the bunched peak at short delay. This time-resolved PL (black curve) represents the decay of the $X$ photon where the start trigger is the detection of an $XX$ photon collected from the opposite facet. For comparison, we also show the decay obtained using a synchronization pulse from the laser (red curve) as in Fig.~\ref{fig:trad_g2}(b). Using the laser trigger, we observe a bi-exponential decay as before with similar extracted parameters: $t_{b,X}= 1.765\pm0.003$\,ns, $t_{d,X}=16.2\pm0.3$\,ns and $A_d/A_b=0.036$. Triggering on the $XX$ photon, we observe a mono-exponential decay with an extracted time constant of $t_b= 1.442\pm0.005$\,ns, over 300\,ps faster. Since detection of the $XX$ photon indicates that the system is in the bright $X$ state, the slow decay process, associated with a transition from bright to dark via a spin flip at a rate $1/t_{d,X}$, will never be observed. The 300\,ps shorter decay extracted from the cross-correlation trace likely represents a more accurate measure of the radiative lifetime since it is not polluted by excitation and spin-flip processes.

\section{Conclusion}

We describe a versatile hybrid platform for efficiently generating single photons on chip based on III-V nanowire quantum dots integrated with Si$_3$N$_4$ photonic circuitry. The approach utilizes a nanomanipulator to place individual nanowires on curved Si$_3$N$_4$ waveguides providing evanescent coupling of single photons emitted from both ends of the nanowire. We use the bi-directional collection to demonstrate single photon emission without the need of an external beamsplitter and to perform cross-correlation measurements for accurate determination of the exciton radiative lifetime. 

The platform paves the way for designing scalable architectures that can efficiently harvest $> 90$\% of the photons emitted from nanowire quantum dots. Such architectures are useful for the study of coherent light-matter interactions with solid-state two-level systems. They will also be required in future quantum information processing technologies based on the availability of multiple on-chip sources of high coherent single photons.

\section*{Acknowledgments}

This work was supported by the Natural Sciences and Engineering Research Council of Canada through the Discovery Grant SNQLS and by the National Research Council of Canada through the Small Teams Ideation Program QPIC. 

\section*{DATA AVAILABILITY}
Data supporting the findings of this study are available from the corresponding author upon reasonable request.

\bibliography{whiskers}   

\end{document}